\RequirePackage{ifpdf}
\ifpdf % We are running pdfTeX in pdf mode
\documentclass[pdftex]{sigma}
\else
\documentclass{sigma}
\fi

\def\sg{\sigma}

\def\lm{\lambda}

\newcommand\Om\Omega

\newcommand\minus\backslash

\def\1{\'{\i}}

\def\>#1{{\mathbf#1}}

\begin{document}

\allowdisplaybreaks

\renewcommand{\thefootnote}{$\star$}

\renewcommand{\PaperNumber}{012}

\FirstPageHeading

\ShortArticleName{B\"acklund Transformations for the Trigonometric Gaudin Magnet}

\ArticleName{B\"acklund Transformations\\ for the Trigonometric Gaudin Magnet\footnote{This paper is a contribution to the Proceedings of the XVIIIth International Colloquium on Integrable Systems and Quantum Symmetries (June 18--20, 2009, Prague, Czech Republic).  The full collection is
available at
\href{http://www.emis.de/journals/SIGMA/ISQS2009.html}{http://www.emis.de/journals/SIGMA/ISQS2009.html}}}

\Author{Orlando RAGNISCO and Federico ZULLO}

\AuthorNameForHeading{O.~Ragnisco and F.~Zullo}

\Address{Dipartimento di Fisica Universit\'a Roma Tre and Istituto Nazionale di Fisica Nucleare,\\ Sezione di Roma, I-00146 Roma, Italy}
\Email{\href{mailto:ragnisco@fis.uniroma3.it}{ragnisco@fis.uniroma3.it}, \href{mailto:zullo@fis.uniroma3.it}{zullo@fis.uniroma3.it}}
\URLaddress{\url{http://webusers.fis.uniroma3/~ragnisco/}}

\ArticleDates{Received December 12, 2009, in f\/inal form January 27, 2010;  Published online January 29, 2010}

\Abstract{We construct a B\"acklund transformation for the trigonometric classical
Gaudin magnet starting from the Lax representation of the model. The Darboux
dressing matrix obtained depends just  on one set of variables because of the
so-called \emph{spectrality} property introduced by E.~Sklyanin and
V.~Kuznetsov. In the  end we mention some possibly interesting open problems.}


\Classification{37J35; 70H06; 70H15}

\section{Introduction}
B\"acklund transformations are a prominent tool in the theory of integrable
systems and soliton theory. Historically they appeared f\/irst in the works of
Bianchi \cite{Bianchi} and B\"acklund \cite{Backlund} on  surfaces of constant curvature and allowed them
to pass from a surface of constant curvature to a new one, or from a solution
of a given PDE to a new one. By this point of view B\"acklund transformations
have been extensively exploited \cite{Rogers,Matveev,Adler M,Levi}. In the f\/ield of
f\/inite-dimensional systems they can be seen as  integrable Poisson maps that
discretize a family of continuous f\/lows; one of the earliest account of this
subject is in~\cite{Ves} where the term \emph{integrable  Lagrange correspondences} is used for \emph{integrable maps}.
This point of view has been widely explored by Suris~\cite{Sur2}, Sklyanin~\cite{S1}, Sklyanin and Kuznetsov~\cite{SK},
Kuznetsov and Vanhaecke~\cite{KV}. Numerous relevant results appeared in the 90's and at the beginning of the
present century on exact time discretizations of many body systems.
Our paper is an ideal continuation, almost 10 years later, of a joint paper by
our dear friend Vadim, Andy Hone and O.R.~\cite{HKR}, where the same problem has
been studied and solved for the rational Gaudin chain.
The key observation we make (see also~\cite{Hik}) is that the {\it trigonometric} Gaudin model with
$N$ sites is just the {\it rational} Gaudin model with $2N$ sites with an
extra {\it reflection} symmetry (``inner automorphism''), entailing the following involution on the corresponding Lax matrix:
\begin{gather}
L (z) = \sigma_3 L (-z) \sigma_3,\label{Symm}
\end{gather}
where $z$ is the spectral parameter, and $\sigma_3$ is the usual Pauli matrix
${\rm diag}\, (1,-1)$. In the following section we will derive (\ref{Symm}) from the
standard form of the trigonometric Lax matrix. Here we can already argue that,
to preserve the ref\/lection symmetry, the elementary dressing matrix, that
we will call $D$ after Darboux, has to enjoy a similar property (up to an
inessential scalar factor), and consequently it has to exhibit pairs of
singular points in the spectral complex plane. Those singular points can be
(opposite) zeroes and/or  (opposite) poles, due to the symmetric role played
by $D$ and $D^{-1}$. As the B\"acklund transformation between the ``old'' Lax
matrix $L$ and the updated $\tilde L$ has to preserve the spectral invariants
of $L$,  it  has to be  def\/ined through a~similarity map:
\begin{gather}
\tilde L (z) = D(z) L(z)[D(z)]^{-1}.\label{BT}
\end{gather}
Obviously we should require that the rational structure of the Lax matrix be preserved, i.e.\ that the updated matrix has  the same number of poles and zeroes as  the old one.
In the sequel we will focus our attention on {\it elementary} B\"acklund transformations, where the corresponding $D$ has just one pair of (opposite) singular points.

\section{The trigonometric Gaudin magnet}

As it is well known the trigonometric Gaudin model is governed by the following
Lax matrix:
\begin{gather}\label{eq:lax}
L(\lm) =   \left( \begin{array}{cc} A(\lm) & B(\lm)\\C(\lm)&-A(\lm)\end{array}
\right),
\\
\label{ABC}
A(\lm)=\sum_{j=1}^{N}\cot(\lm-\lm_{j})s^{3}_{j}, \qquad
B(\lm)=\sum_{j=1}^{N}\frac{s^{-}_{j}}{\sin(\lm-\lm_{j})},\qquad C(\lm)=\sum_{j=1}^{N}\frac{s^{+}_{j}}{\sin(\lm-\lm_{j})}.
\end{gather}
The dynamical variables $\big(s^{+}_{j},s^{-}_{j},s^{3}_{j}\big)$, $j=1, \ldots, N$,
obey to the Poisson structure given by the brackets:
\begin{gather*}
\big\{s^{3}_{j},s^{\pm}_{j}\big\}=\mp i\delta_{jk}s^{\pm}_{k}, \qquad
\big\{s^{+}_{j},s^{-}_{j}\big\}=-2i\delta_{jk}s^{3}_{k},
\end{gather*}
with the $N$ Casimirs given by
\begin{gather*}
s_{j}^{2}=\big(s_{j}^{3}\big)^{2}+s_{j}^{+}s_{j}^{-}.
\end{gather*}
This structure corresponds to the trigonometric $r_{t}$ matrix, given by
\begin{gather*}
r_{t}(\lm) = \frac{1}{\sin(\lm)}\left(\begin{array}{cccc} \cos(\lm)& 0 & 0 &0 \\
0 & 0 & 1 & 0\\
0 & 1 & 0 & 0\\
0 & 0 & 0 & \cos(\lm)
\end{array}\right),
\end{gather*}
with the Lax matrix satisfying the \emph{linear} $r$-matrix Poisson algebra,
\begin{gather} \label{eq:pois}
\big\{L^{1}(\lm),L^{2}(\mu)\big\}=\big[r_{t}(\lm-\mu),L^{1}(\lm)+L^{2}(\mu)\big],
\end{gather}
where, as usually, the superscripts on the matrices denote tensor products:
\[
L^{1}=L\otimes I, \qquad L^{2}=I\otimes L.
\]
The equation (\ref{eq:pois}) is equivalent to the following Poisson brackets
for the elements $A(u)$, $B(u)$ and~$C(u)$:
\begin{gather*}
 \{A(\lm),A(\mu)\}=\{B(\lm),B(\mu)\}=\{C(\lm),C(\mu)\}=0,\nonumber\\
 \{A(\lm),B(\mu)\}=\frac{\cos(\lm-\mu)B(\mu)-B(\lm)}{\sin(\lm-\mu)},\nonumber\\
 \{A(\lm),C(\mu)\}=\frac{C(\lm)-\cos(\lm-\mu)C(\mu)}{\sin(\lm-\mu)},\nonumber\\
 \{B(\lm),C(\mu)\}=\frac{2(A(\mu)-A(\lm))}{\sin(\lm-\mu)}.%\label{eq:aabbcc}
\end{gather*}
Through the ``uniformization'' mapping:
\begin{gather*}
\lm \to z =e^{i\lm} %\label{uni}
\end{gather*}
the Lax matrix (\ref{eq:lax}) acquires a rational form in $z$:
\begin{gather}
\label{eq:ratiL}
-iL(z)=\sum_{j=1}^{N}s^{3}_{j}\sigma_3+\sum_{j=1}^{N}
\left(\frac{L_{1}^{j}}{z-z_{j}}-\sigma_3\frac{L_{1}^{j}}{z+z_{j}}\sigma_3\right),
\end{gather}
where the matrices $L_{1}^{j}$, $j=1, \ldots, N$, have the simple form:
\begin{gather*}
L_{1}^{j}=z_{j}\left(\begin{array}{cc}s_{j}^{3} & s_{j}^{-}\vspace{1mm}\\
s_{j}^{+} & -s_{j}^{3}\end{array}\right).
\end{gather*}
The equation (\ref{eq:ratiL}) leads to the ref\/lection symmetry (\ref{Symm}):
\[
L(z)=\sg_{3}L(-z)\sg_{3}.
\]

\section{The Darboux matrix}
The simplest  choice for the spectral structure of the Darboux-dressing
matrix requires that it obeys the ref\/lection symmetry (\ref{Symm}) and contains only one
pair of opposite simple poles. Then, it reads:
\begin{gather}
D = D_\infty + \frac{D_1}{z - \xi}-\frac{\sigma_3 D_1\sigma_3}{z +\xi}.\label{first}
\end{gather}
The  matrix $D_\infty$, i.e.\ $\lim\limits_{z\to \infty} D(z)$ def\/ines the
\textit{normalization} of the problem. The equation (\ref{BT}), rewritten in the form:
 \begin{gather}
\tilde L (z) D(z) = D(z) L(z)\label{BTnew}
\end{gather}
in the limit $z \to \infty$ yields:
\begin{gather*}
(\tilde S_z)\sigma_3D_\infty = D_\infty (S_z)\sigma_3,
\end{gather*}
where by $S_z$ we have denoted the $z$-component of the total ``spin'' $S$.
As $S_z$ Poisson commutes with ${\rm tr}\, L^2$, the generating function of the
complete family of involutive Hamiltonians, it has  to  be preserved by
our B\"acklund transformation, which is a symmetry for the whole
hierarchy. This implies $D_\infty$ to be diagonal.

As for bounded values of $z$,  equation \eqref{BTnew} implies  that both sides have equal residues at the simple poles $ \pm z_{j}$, $\pm \xi$. However, in view of the symmetry property (\ref{Symm}), (\ref{first}) it will be enough to look at half of them, say $ z_{j}$, $\xi$.
The corresponding equations will be:
\begin{gather}
\tilde L_{1}^{(j)} D(z_{j}) = D(z_{j}) L_{1}^{(j)}, \label{resj}
\\
\tilde L (\xi) D_{1} = D_{1} L(\xi). \label {resxi}
\end{gather}
The crucial problem to solve now is to ensure that (\ref{resj}), (\ref{resxi}) provide an explicit (and symplectic) mapping between the old and the new spin variables. In other words, to get a Darboux matrix that depends {\it just} on one set of variables, say the old ones.
As it has been shown for instance in~\cite{SK,HKR}, this can be done thanks to the so-called {\it spectrality} property.
In the present context, this amounts to require that $\det D$ possess, in
addition to the two opposite  poles $\pm\xi $, two opposite {\it nondynamical}
zeroes, say $\pm \eta$ and that $D_{1}$ is, up to a factor, a projector. Again, by symmetry it will be enough to look at one of the zeroes, say~$\eta$.
By setting $z=\eta$ in (\ref{BTnew}) we get
\begin{gather*}
\tilde L (\eta) D(\eta) = D(\eta) L(\eta).
\end{gather*}
But $D(\eta)$ is a rank one matrix, having a one dimensional Kernel $|K(\eta)\rangle$, whence:
\begin{gather*}
0 = D(\eta) L(\eta)|K(\eta)\rangle
\end{gather*}
entailing
\begin{gather}
\label{spectral1}
 L(\eta)|K(\eta)\rangle = \mu (\eta) |K(\eta)\rangle,
\end{gather}
 i.e. the points $\pm \eta$, $\pm \mu(\eta)$ belong to the \textit{spectral curve} $\det (L(z) -\mu I) =0$. $|K(\eta)\rangle$ is then fully determined in terms of the old dynamical variables.
The equation (\ref{resxi}) give us another one dimensional Kernel~$K(\xi)$
because also $D_{1}$ is a rank~1 matrix, so (\ref{resxi}) entails:
\begin{gather}
\label{spectral2}
 L(\xi)|\Omega(\xi)\rangle = \mu (\xi) |\Omega(\xi)\rangle.
\end{gather}
The two spectrality conditions (\ref{spectral1}), (\ref{spectral2}) allow to
write $D$ in terms of the old dynamical variables and of the two B\"acklund
parameters $\xi$ and $\eta$, so that the B\"acklund equations (\ref{resj})
yield an explicit map between the  new (tilded) and the old (untilded) dynamical variables.
In order to clarify the point above let us make some observations.
\noindent First of all note that requiring $D_{1}$ to  be a rank one matrix amounts to require that
the determinant of $(z^{2}-\xi^{2})D(z)$ be zero for $z=\xi$ or, by symmetry,
for $z=-\xi$. In fact:
\begin{gather*}
\big(z^{2}-\xi^{2}\big)D(z)|_{z=\xi}=2\xi D_{1}, \qquad \big(z^{2}-\xi^{2}\big)D(z)|_{z=-\xi}=2\xi\sigma_3D_{1}\sigma_3.
\end{gather*}
Since  two Darboux matrices dif\/fering just by a multiplicative scalar factor
def\/ine the same BT, we can choose to work with a modif\/ied Darboux matrix $D'(z)$ def\/ined by the relation:
\begin{gather*}
%\label{eq:d'}
D'(z)\equiv \frac{z^{2}-\xi^{2}}{z}D(z).
\end{gather*}
Hence, to ensure that the spectrality property holds true we have to require
 $\det  D'(z)$ to vanish at $z=\xi$ and $z=\eta$. The form
taken by  the Darboux matrix $D'(z)$ can be further simplif\/ied by writing:
\begin{gather}
D'(z) = z^{-1}\hat{A} + \hat{B} + \hat{C}z.\label{second}
\end{gather}
The matrix $\hat{C}$ is immediately seen to be a diagonal one by looking at the behavior for large values of~$z$ and requiring $\tilde S_z = S_z$.
On the other hand,  $L(0)$ as well as its dressed version $\tilde L(0)$ are diagonal matrices:
\begin{gather*}
L(0) =  \sum_{j=1}^N S_z^{(j)}\sigma_3-\sum_{j=1}^N\frac{L_1^{(j)}+\sigma_3L_1^{(j)}\sigma_3}{z_j}.
\end{gather*}
This readily implies that $\hat{A}$ in (\ref{second}) is diagonal.
In turn,  (\ref{Symm}) implies that if  even powers of~$z$ are
diagonal, odd powers must be of\/f-diagonal, entailing that~$\hat{B}$ is an of\/f-diagonal matrix.
The two matrices $\hat{A}$ and $\hat{C}$ are then given respectively by
${\rm diag}\,(a_{1},a_{2})$, ${\rm diag}\,(c_{1},c_{2})$, whereas the of\/f-diagonal matrix
$\hat{B}$ is given by $\left(\begin{array}{cc}0&b_{1}\\b_{2}&0 \end{array}\right)$.

We get a deeper insight on the parameterization of matrices $\hat{A}$, $\hat{B}$, $\hat{C}$ resorting
again to the spectrality property: we stress once more that this amounts to
requiring  $D'(\xi)$ and $D'(\eta)$ to be  rank one matrices. This means
that there exists a  function of one variable,  say~$p$, such that:
\begin{gather}
\left\{\begin{aligned}
&c_1\xi +a_1/\xi +b_1p(\xi) =0,\\
&b_2 + p(\xi)(c_2\xi + a_2/\xi)=0;\end{aligned}\right.\label{p}
\\
\left\{\begin{aligned}
&c_1\eta +a_1/\eta +b_2p(\eta)=0,\\
&b_2 + p(\eta)(c_2\eta + a_2/\eta)=0.\end{aligned}\right.\label{q}
\end{gather}
The four equations (\ref{p}), (\ref{q}) leave us with two undetermined
parameters, one of which is a~global multiplicative factor for $D'(z)$, say
$\beta$. The other is denoted by $\gamma$. The parameterization of $D'$ reads as follows:
\begin{gather*}
%\label{eq:D}
D'(z)= \beta\left( \begin{array}{cc} {\frac {z \left( p(\eta)\eta-p(\xi)\xi \right)
    }{\gamma}}+{\frac{ \left( p(\xi)\eta-p(\eta)\xi \right) \eta\xi}{\gamma z}}&{\frac {{\xi}^{2}-{\eta}^{2}}{\gamma}}\vspace{2mm}\\
 {\frac {\gamma p(\xi)p(\eta) \left(
      {\xi}^{2}-{\eta}^{2} \right) }{\eta\xi}}&{\frac {\gamma \left( p(\eta)\eta-p(\xi)\xi \right)
    }{z}}+{\frac {  \gamma z\left( p(\xi)\eta-p(\eta)\xi \right)}{\eta\xi}}\end{array} \right).
\end{gather*}
The kernel of $D(\xi)$ (resp.\ $D(\eta))$) is simply given by the row
$|\Omega(\xi)\rangle =(1,p(\xi))^{T}$ (resp.\ $|\Omega(\eta)\rangle =(1,p(\eta))^{T}$). It is an   eigenvectors of $L(\xi)$ (resp.\ $L(\eta)$).
Hence  $p(\xi)$  can be written as:
\begin{gather*}
p(\xi)=\frac{\mu(\xi)-A(\xi)}{B(\xi)},
\end{gather*}
where we recall that $\mu(z)$ is such that $\mu^{2}(z)=A^{2}(z)+B(z)C(z)$ and
$A(z)$, $B(z)$, $C(z)$ are given by (\ref{ABC}). In terms of  $p(\eta)$, $p(\xi)$, the matrices $D_{\infty}$ and $D_{1}$ in (\ref{first}) take the
form:
\begin{gather*}
D_{\infty}=\beta\left(\begin{array}{cc}\frac{p(\eta)\eta-p(\xi)\xi}{\gamma}&0
\vspace{1mm}\\0&\gamma\frac{p(\xi)\eta-p(\eta)\xi}{\eta
\xi} \end{array}\right),
\\
D_{1}=\beta\big(\eta^{2}-\xi^{2}\big)\left(\begin{array}{cc}\frac{p(\xi)}{\gamma}&-\frac{1}{\gamma}
\vspace{1mm}\\
-\gamma\frac{p(\xi)p(\eta)}{\eta
\xi}& \gamma\frac{p(\eta)}{\eta\xi}\end{array}\right).
\end{gather*}
Since the Darboux matrix $D(z)$ is completely known in terms of one set of dynamical
variables, equation (\ref{resj}) yields an explicit B\"acklund transformation
for the trigonometric Gaudin magnet.

In a forthcoming paper we will prove that (\ref{resj}) provides indeed a symplectic
map between old and new dynamical variables, and moreover that, according to a
Sklyanin conjecture, the Darboux matrix (\ref{second}) is in fact identical to
Lax matrix of the elementary trigonometric Heisenberg magnet. The interpolating Hamiltonian f\/low will be also derived and some examples of discrete dynamics will be displayed and discussed.

\subsection*{Acknowledgments}

This paper is intended to be a contribution to the Proceedings of the International Conference ``Integrable Systems and Quantum Symmetries 2009'', organized by Professor \v{C}.~Burd\'{\i}k and held in Prague, June 18--20, 2009.
One of the authors (O.R.) wants to warmly thank for his hospitality the Newton Institute for Mathematical Sciences, and all the organizers and the participants to the Program ``Discrete Integrable Systems''. It was in fact during his stay in Cambridge that   the main  ideas presented in the paper have been made precise. Also, he acknowledges enlightening discussions with A.~Levine (ITEF) at the workshop  ``Einstein at SISSA 2009'', partially funded by the Russian Foundation for Basic Research within the project ``The Theory of Nonlinear Integrable Systems''.

\pdfbookmark[1]{References}{ref}
\LastPageEnding

\end{document}